\documentclass[conference, a4paper]{IEEEtran}
\IEEEoverridecommandlockouts                        
\usepackage{multirow, makecell}
\usepackage{amsmath,amssymb,latexsym}
\usepackage{rotating}
\usepackage{lettrine}
\usepackage{textcomp}
\usepackage{bm}

\usepackage{mathtools}
\usepackage{siunitx}

\usepackage[table]{xcolor}
\usepackage{hyperref}
\usepackage{tikz}

\usepackage{xcolor}

\usepackage{lipsum}
\usepackage[pscoord]{eso-pic}
\newcommand{\placetextbox}[3]{
  \setbox0=\hbox{#3}
  \AddToShipoutPictureFG*{
    \put(\LenToUnit{#1\paperwidth},\LenToUnit{#2\paperheight}){\vtop{{\null}\makebox[0pt][c]{#3}}}%
  }%
}%

\newcommand{\SM}[1]{\widehat{\mathcal{S}}^{M_{#1}}}

\title{Widening the Coverage of Reference Broadcast Infrastructure Synchronization in Wi-Fi Networks 
\thanks{This work was partially supported by the European Union under the Italian National Recovery and Resilience Plan (NRRP) of NextGenerationEU, partnership on ``Telecommunications of the Future'' (PE00000001 - program ``RESTART'').}
}

\author{
    \IEEEauthorblockN{
    Gianluca Cena\IEEEauthorrefmark{1},
    Pietro Chiavassa\IEEEauthorrefmark{1},
    Gabriele Formis\IEEEauthorrefmark{1}\IEEEauthorrefmark{2},
    Stefano Scanzio\IEEEauthorrefmark{1},
    }  
    \IEEEauthorblockA{\IEEEauthorrefmark{1}National Research Council of Italy (CNR--IEIIT), Italy. \IEEEauthorrefmark{2}Politecnico di Torino, Italy.}    
    Email:  \{gianluca.cena, pietrochiavassa, stefano.scanzio\}@cnr.it, \{gabriele.formis\}@polito.it
}

\begin{document}
\placetextbox{0.5}{1}{This is the author's version of an article that has been published.}
\placetextbox{0.5}{0.985}{Changes were made to this version by the publisher prior to publication.}
\placetextbox{0.5}{0.97}{The final version of record is available at \href{https://doi.org/10.1109/WFCS63373.2025.11077631}{https://doi.org/10.1109/WFCS63373.2025.11077631}}%
\placetextbox{0.5}{0.05}{Copyright (c) 2025 IEEE. Personal use is permitted.}
\placetextbox{0.5}{0.035}{For any other purposes, permission must be obtained from the IEEE by emailing pubs-permissions@ieee.org.}%

\maketitle
\thispagestyle{empty}
\pagestyle{empty}

\begin{abstract}
Precise clock synchronization protocols are increasingly used to ensure that all the nodes in a network share the very same time base.
They enable several mechanisms aimed at improving determinism at both the application and communication levels, which makes them highly relevant to industrial environments.
Reference Broadcast Infrastructure Synchronization (RBIS) is a solution specifically conceived for Wi-Fi that exploits existing beacons and can run on commercial devices.

In this paper, an evolution of RBIS is presented, we call DOMINO, whose coverage area is much larger than the single Wi-Fi infrastructure network, potentially including the whole plant.
In particular, wireless stations that can see more than one access point at the same time behave as boundary clocks and propagate the reference time across overlapping networks.
\end{abstract}

\section{Introduction}
\label{sec:intro}
Clock synchronization protocols (CSP) are a peculiar class of network protocols that is becoming increasingly relevant, particularly in industrial environments \cite{SynchronizeYourWatches2013_I, SynchronizeYourWatches2013_II}.
Unlike communication protocols, CSPs are not aimed to support information exchanges among end nodes.
Instead, they serve to provide nodes with the same (shared) time base.
It is worth noting that the same goal can be also obtained by other means.
For example, Global Navigation Satellite Systems (GNSS), 
which include the Global Positioning System (GPS), Global Navigation Satellite System (GLONASS), BeiDou Navigation Satellite System (BDS), and Galileo, 
permit nodes to determine both their position and a precise estimate of the current Coordinated Universal Time (UTC).
However, GNSSs suffer from several limitations: 
1)~Satellite coverage may be problematic in certain areas, especially indoor (factory shop floors);
2)~Specific hardware is needed, which increases both costs and energy consumption;
3)~The countries that own satellites may decide to stop/restrict their usage at any time.

The most known CSP is probably the Network Time Protocol (NTP) \cite{rfc5905_NTP}, whose definition dates back to 1985.
NTP relies on UDP/IP (hence supporting a variety of underlying physical networks)
and is currently the standard solution to synchronize computers over the Internet.
The synchronization error of NTP may exceed \SI{1}{ms} \cite{2019_ISPCS}, and can be quite large even on local area networks, which makes it unsuitable for many industrial control applications.
For this reason, when precise and accurate time synchronization is required, other solutions are needed.
The Precision Time Protocol (PTP), also known as IEEE 1588 \cite{2019_IEEE1588}, was purposely defined for industrial networks and achieves sub-\SI{}{\mu s} accuracy.
An adaptation of PTP, called generalized PTP (gPTP), constitutes the standard solution in the context of Time-Sensitive Networking (TSN), where it is known as IEEE 802.1AS
\cite{2020_IEEE802.1AS}.

Although PTP-like solutions can operate on wireless networks as well, and in particular on IEEE 802.11 (Wi-Fi) \cite{IEEE802.11-20}, as a matter of fact at present they are mainly employed in IEEE 802.3 (Ethernet) networks.
In Wi-Fi, a time synchronization function (TSF) is envisaged that consists of a $\SI{64}{b}$ counter that is increased once per $\SI{}{\mu s}$ and is included by the access point (AP) in beacons.
Upon reception, every associated wireless station (STA) updates its local TSF counter with the TSF value included in the beacon.
As explained in \cite{2017-TII-Mahmood}, only offset correction is carried out (and not rate correction), which makes the TSF mechanism suffer from time warps and poor precision (error may be as high as some tens $\SI{}{\mu s}$).

Since the version of IEEE 802.11 dated 2012, a time advertisement (TA) tag can be included in beacon frames as an information element, which relates the TSF counter with the local time of the AP itself (e.g., the offset between them).
This means that, in theory, the local clock of a STA can be precisely aligned to the local clock of the AP.
However, it is unclear from the standard how the local clock and the TSF counter of the AP can be practically synchronized. 
Moreover, according to \cite{2017-TII-Mahmood}, chipsets capable to include the time offset in beacon frames on-the-fly are hardly available off the shelf.

In \cite{2021-ICIT} another sender-receiver clock synchronization mechanisms is described.
The AP takes a precise timestamp on beacon transmission, which is subsequently distributed using a \textit{Follow\_Up} message.
Every STA associated to the AP can then adjust its local clock, so that it coincides with the time base of the AP.
This technique can run on recent commercial Wi-Fi boards, but requires the firmware of the AP to be modified.
Since 2016, Wi-Fi specifications include a fine-time measurement (FTM) mechanism that can be used by STAs to evaluate the round-trip time (RTT) to the AP, and hence their distance from it.
This mechanism is mostly exploited for device localization, but it could serve for precise clock synchronization as well,
by measuring the propagation delay.
Only a limited number of chipsets support this function.

When mesh wireless networks are taken into account, an inexpensive alternative for clock synchronization is the Reference Broadcast Synchronization (RBS) \cite{2003_RBS}.
In this case, specific messages are exploited as common synchronization events (\textit{Sync}), on the reception of which all nodes take precise timestamps (receiver/receiver paradigm).
This can be done easily, since the wireless spectrum (over limited areas) can be thought as a broadcast domain.
These timestamps are then exchanged among nodes, so that a shared time base can be determined to which local clocks are adjusted.

An interesting variation of the above mechanism is found in the Reference Broadcast Infrastructure Synchronization (RBIS) \cite{2012-ETFA-RBIS, ImplementationEvaluationReference2015},
which was purposely conceived for infrastructure Wi-Fi networks (characterized by an AP to which all STAs associate).
Each such network is denoted as a Basic Service Set (BSS).
In this case, beacons sent by the AP are taken as the \textit{Sync} events by STAs.
This brings two advantages: 
1) STAs do not have to worry about \textit{Sync} generation, and, 
2) network traffic caused by the CSP is lowered, since the beacons it relies on are generated by the AP irrespective of it.
An implementation of RBIS is currently being developed as open source \cite{2024-ETFA}.
Moreover, the RBIS idea has been recently proposed for achieving time synchronization in 5G networks as well \cite{2022-GCWkshps}.
The main problem of RBIS is that, in its current version, it can only synchronize STAs within the same Basic Service Set (i.e., the STAs associated to the same AP).
This is a non-negligible limitation when large environments are involved, like industrial plants.

In this paper an extension of RBIS is proposed, we term DOMINO, which potentially achieves much wider coverage.
It combines concepts from RBIS, PTP, and multi-hop RBS, by permitting STAs that observe beacons from more than one AP to act as boundary clocks.
The real advantage of DOMINO is that it has been conceived to enable simple implementations on commercial Wi-Fi boards.
The basic idea is not dramatically different from the mechanism proposed in \cite{2021_ICPS}, where the offset between beacons sent by different APs was evaluated by STAs and used to maintain a centralized database.
However, our solution is aimed at providing high reliability in very dynamic environments, where STAs keep on moving around and the likelihood that a sizable amount of beacons is lost can not be neglected.

The paper is structured as follows:
in Section~\ref{sec:RBIS} RBIS is briefly described and some hints on how its coverage can be enlarged are given.
In Section~\ref{sec:DOMINO} the new DOMINO protocol is presented.
Finally, in Section~\ref{sec:concl}, some conclusions are drawn.

\section{Widening RBIS Coverage}
\label{sec:RBIS}
The spectrum available for Wi-Fi communication is limited, which means that in densely populated areas BSSs almost unavoidably overlap, at least in part.
This is particularly true for the \SI{2.4}{GHz} band, which is still in widespread use because it is virtually supported by every commercial adapter.
As long as APs operating on the same channel (or on partially overlapping channels) are not too close to each other, this does not preclude satisfactory communication quality.

By a simple extension of the RBIS protocol, this situation can be leveraged to synchronize the local clocks of the STAs deployed on a wide area without having to bring any changes to the existing network infrastructure.
In fact, unlike solutions that rely on time-sensitive networking (TSN) like IEEE 802.1AS
(which demands compliant Ethernet switches), 
no modifications are required by RBIS (and DOMINO) to APs, but only to STAs.
To support legacy \mbox{Wi-Fi} boards, features like FTM to compute and compensate propagation delays will not be considered, even though they can be optionally exploited by those STAs supporting them.

\subsection{RBIS Protocol Basics}
The basic RBIS protocol \cite{2012-ETFA-RBIS, ImplementationEvaluationReference2015} relies on concepts borrowed from both PTP and RBS.
Its operation is quite simple, as it resembles the PTP synchronization mechanism (not including propagation delay calculation, about \SI{3}{ns/m}).
Unlike PTP, however, RBIS exploits beacon receptions as the common events on which the STAs in the BSS that want to synchronize their clocks take precise timestamps.
Hence, there is no need for explicit \textit{Sync} messages, which results in lower communication overheads.
RBIS is a receiver-receiver CSP, i.e., it relies only on timestamps acquired on received packets. 
For this reason, it is easier to implement than PTP. 

\begin{figure}[b]
    \centering
    \includegraphics[width=1\columnwidth]{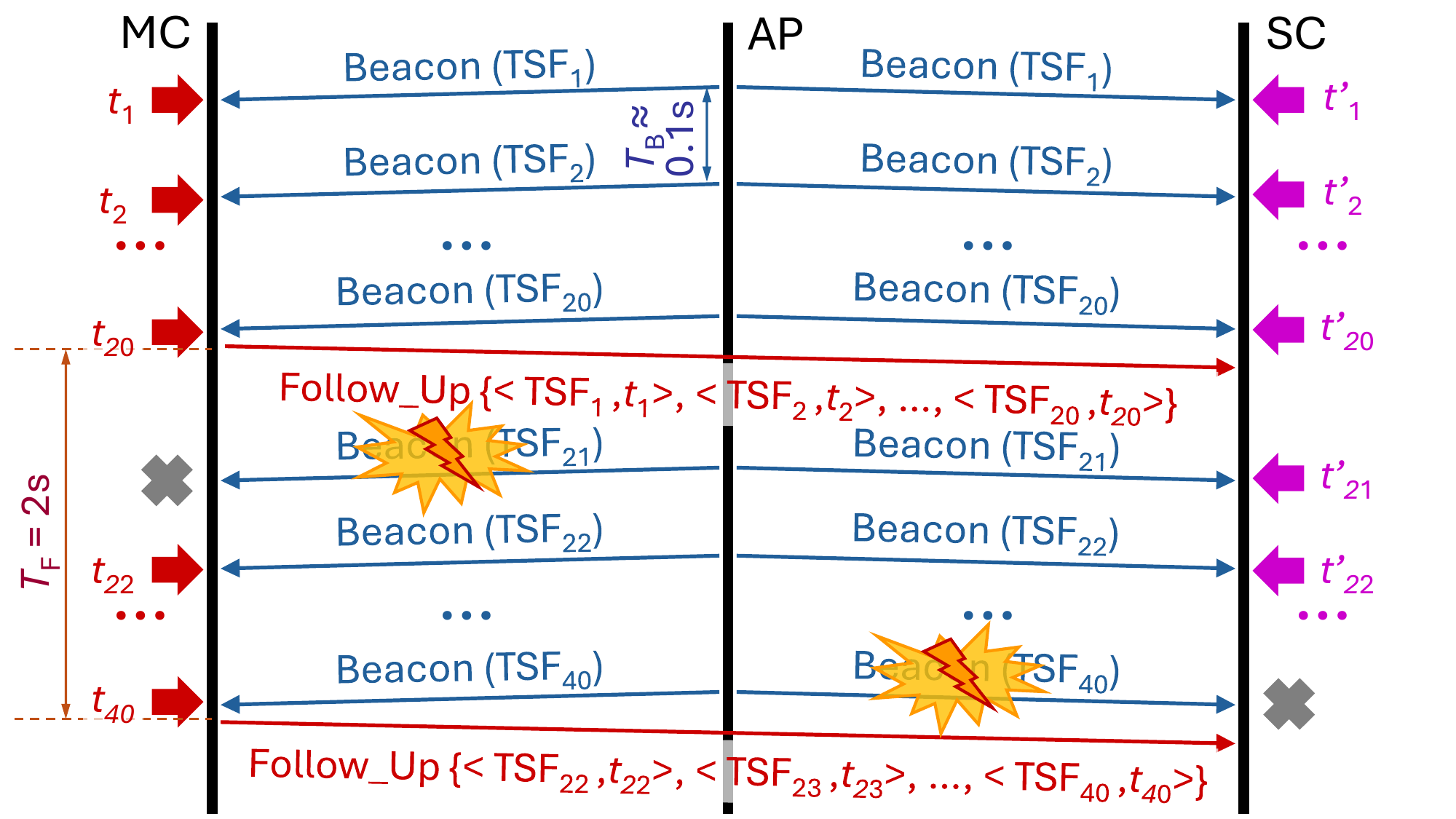}
    \caption{Basic RBIS operation (some beacons are lost).}
    \label{fig:basic}
\end{figure}

One of the STAs in the BSS, known as the \textit{master clock} (MC) and denoted $M$, repeatedly broadcasts a \textit{Follow\_Up} (FUP) message that includes its timing information about the most recent beacons it received.
Embedding details about more than one beacon increases robustness against frame losses, an aspect that should not be neglected for broadcast transmissions.
Such information is encoded as a sequence of pairs $\langle \mathrm{TSF}_k, t_k \rangle$ (one per beacon),
where $\mathrm{TSF}_k$ is the value of the TSF field included in the $k$-th beacon frame (expressed in microseconds) 
while $t_{k}$ is a precise timestamp taken by $M$ on its arrival.
The period of FUP messages must be strictly longer than the beacon period,
and is typically selected to be several times larger (e.g., $\SI{2}{s}$) when multiple beacons are included.
Operation of RBIS is sketched in Fig.~\ref{fig:basic}.

While the TSF made available by commercial equipment is customarily considered not precise enough (error may exceed ten $\SI{}{\mu s}$ \cite{2014_WFCS_TSF}), the timestamp offered by recent Wi-Fi adapters can be much more precise, with a resolution of $\SI{10}{ns}$.
The idea is to use the TSF value as a unique identifier for beacons, so that every \textit{slave clock} (SC) can pair its precise timestamps $t'_{\kappa}$ with those made available in the FUP message.
As can be seen from Fig.~\ref{fig:basic}, unless losses are particularly severe at least one match is typically found, which enables synchronization.
Once timestamps have been paired, a suitable clock discipline algorithm (CDA) running in every STA is employed to synchronize its local clock to the clock of the master. 
In this way a shared time base is provided network-wide that enables synchronous operations of STAs, 
e.g., coordinated medium access or distributed real-time control applications.
In the case of software timestamps, the synchronization error of RBIS is generally larger, but it can be kept well below $\SI{1}{\mu s}$ by using suitable CDAs \cite{7301461}. 
In particular, CDAs based on neural networks can increase accuracy further \cite{8402342}.

\subsection{Definitions}
Some definitions are now given to describe the mechanism that permits to enlarge the coverage area of RBIS.

Let  $\mathcal{A} \doteq \{A_i, i\in[1,n] \}$ be the set of all the APs deployed in the area of interest, where $A_i$ denotes a generic AP.
We will assume that they are quite near or, better, that given any pair of APs either they can hear from each other or one or more paths exist between them traversing a number of intermediate APs, where every such AP can hear from both the previous and the next AP in the path.
This is just to describe in general terms the network topology: APs are conventional ones and so they are not assumed to communicate with each other over the air.
As customarily happens in the real world, APs are interconnected through a wired network.
In particular, in the following we will assume that all APs in $\mathcal{A}$ are connected to the same (switched) Ethernet network.

For every AP an infrastructure BSS exists, to which a number of STAs can be associated.
STAs that support clock synchronization as per RBIS (or its extensions) are denoted \textit{time-synchronized} STAs (TS STA).
Note that no TS APs exist, as no modifications are required for them by RBIS. 
Let $S_l$ denote a generic TS STA, associated to some AP $A_i \in \mathcal{A}$, while $\mathcal{S} \doteq \{S_l \}$ represents the set of all TS STAs in the overall network.
Since this paper focuses on TS STAs, for the sake of simplicity in the following they will be often referred to simply as STAs.

Because of the (wired) backbone, all STAs in $\mathcal{S}$ can freely communicate using data-link layer services, including broadcast and multicast frame transmissions.
In other words, $\mathcal{A} \cup \mathcal{S}$ is an Extended Service Set (ESS) and, as such, it constitutes a broadcast domain.
Think to these devices as part of the same Logical IP Subnet (LIS).
Nodes on Ethernet can be also reached, but this is irrelevant to our purposes since they cannot be directly synchronized by RBIS.

Set $\mathcal{S}$ is split in two subsets: 
the STAs that can behave as master clocks, we denote $M_j$, 
collectively referred to as the set $\mathcal{M} \doteq \{M_j, j\in[1,m] \} \subseteq \mathcal{S}$, 
and those that can only behave as slave clocks.
Concerning the CSP, the former are able to play an active role by sending FUP messages, 
whereas the latter are passive subjects that are only able to synchronize their own clock to some master.
To make descriptions about the exchanged messages clearer, in the following we will denote the $\ell$-th FUP message sent by $M_j$ as $\mathfrak{F}^{M_j}_\ell$.
Instead, $\mathfrak{B}^{A_i}_k$ denotes the $k$-th beacon (i.e., \textit{Sync} message) sent by $A_i$.
Subscripts $\ell$ and $k$ will be omitted when not relevant.

Among the MCs, one is selected as the \textit{grandmaster clock} (GC) and provides its local time (which becomes the reference time) to all the other TS STAs (both MCs and SCs).
The GC is often equipped with a GNSS receiver (e.g., a GPS module), so that its local clock is synchronized to the UTC.
As an alternative, it can be connected to the wired network through an Ethernet port, which permits its local clock to be synchronized through PTP or IEEE 802.1AS.

If more than one such nodes are available in $\mathcal{S}$, one of them is dynamically selected as the GC using solutions that resemble the PTP Best Master Clock Algorithm (BMCA).
There are, however, relevant differences due to the peculiar behavior of wireless networks with respect to the wired ones.
In fact, both the position of nodes and the spectrum conditions may keep varying in the former case.
The topic of GC selection is faced only in part in this paper, and its complete definition is left as future work.
For the sake of simplicity, in the following we will assume that GC coincides with $M_1$.

\begin{table}
    \caption{Symbols and notation}
    \label{tab:symbols}
    \centering
    \begin{tabular}{cl}
         Symbol &  Description \\
        \hline
        $S_l \in \mathcal{S}$       & generic time-synchronized STA and set of all TS STAs \\
        $A_i \in \mathcal{A}$       & generic access point and set of all APs  \\
        $M_j \in \mathcal{M}$       & generic master clock and set of all MCs  \\
        $\mathfrak{B}^{A_i}_k$      & $k$-th \textit{Sync} message (beacon) sent by $A_i$ \\
        $\mathfrak{F}^{M_j}_\ell$   & $\ell$-th \textit{Follow\_Up} message (FUP) sent by $M_j$ \\
        $\mathcal{S}^{A_i}$         & set of all STAs that can hear beacons from $A_i$ \\
        $\mathcal{A}^{S_l}$         & set of all APs whose beacons can be heard by $S_l$ \\
        $\SM{j}$                    & set of all STAs that can hear beacons from APs in $\mathcal{A}^{M_j}$ \\
        $A_i \mapsto S_l$           & $S_l$ is associated to $A_i$ \\
        $A_i \rightharpoonup S_l$   & $S_l$ can hear beacons from $A_i$\\
        $M_j \Rightarrow S_l$       & $M_j$ provides SYNOPs to $S_l$ (TS link) \\
        \hline
    \end{tabular}
\end{table}

We use notation $A_i \mapsto S_l$ to indicate that STA $S_l$ (either a MC or a SC) is associated to AP $A_i$.
The infrastructure BSS that hinges on $A_i$ is denoted $\mathrm{BSS}_i = \{ S_l | A_i \mapsto S_l$\},
while $\mathrm{ESS} = \bigcup_{A_i \in \mathcal{A}} \mathrm{BSS}_i$.
To keep notation simple, in this paper APs are not included in sets $\mathrm{BSS}_i$ and $\mathrm{ESS}$.
Similarly, notation $A_i \rightharpoonup S_l$ means that $S_l$ can hear beacons 
from $A_i$.
Condition $A_i \mapsto S_l$ implies that $A_i \rightharpoonup S_l$
(association is terminated by the STA if no beacons are heard for a while), 
but the reverse is not necessarily true since we assumed that BSSs may overlap.

We call the portion of space where beacon frames $\mathfrak{B}^{A_i}_k$ can be heard the \textit{coverage region} of $A_i$.
More precisely, it is defined as the set 
of STAs in $\mathcal{S}$ that can hear beacons from $A_i$, 
that is, $ \mathcal{S}^{A_i} \doteq \{ S_l | A_i \rightharpoonup S_l \}$.
Besides including all STAs associated to $A_i$ (which coincides with $\mathrm{BSS}_i$), 
other nearby STAs associated to neighbor APs can also fit in $\mathcal{S}^{A_i}$,
which implies that $\mathcal{S}^{A_i} \supseteq \mathrm{BSS}_i$
(as will be seen, our hypotheses on devices' placement imply that $\mathcal{S}^{A_i} \supset \mathrm{BSS}_i$).

Let $\mathcal{A}^{S_l}$ be the set of APs whose beacons can be heard by a generic TS STA $S_l$, in formulas, 
$ \mathcal{A}^{S_l} \doteq \{ A_i | A_i \rightharpoonup S_l \}$.
Since every STA must be associated to an AP in order to communicate,
$|\mathcal{A}^{S_l}| \geq 1$ (notation $|\cdot|$ denotes the cardinality of a set).
For MC $M_j$ this set is denoted $\mathcal{A}^{M_j}$.
A necessary condition so that the technique described in this paper can operate correctly is that, at any time, 
a specific subset of $\mathcal{M}$ must hear beacons from at least two distinct APs, that is, $|\mathcal{A}^{M_j}| \geq 2$.
For what said before, this is not a limiting assumption in real scenarios.

Finally, let $\SM{j}$ be the set of all the STAs that can hear at least one of the beacons that are also received by $M_j$, that is, those sent by any of the APs in $\mathcal{A}^{M_j}$, in formulas
\mbox{$\SM{j} \doteq \{ S_l | A_i \in \mathcal{A}^{M_j} \wedge A_i \rightharpoonup S_l\}$.}
By construction, this set includes at least $M_j$, that is, $M_j \in \SM{j}$.
More in general, we have that 
$\SM{j} = \bigcup_{A_i \in \mathcal{A}^{M_j}} \mathcal{S}^{A_i}
\supseteq \bigcup_{A_i \in \mathcal{A}^{M_j}} \mathrm{BSS}_i$.
All STAs in $\SM{j}$ can be time synchronized to $M_j$ through RBIS.
In fact, among the beacons they hear there is always at least one in common with $M_j$, which can be paired upon reception of the related FUP message and used to adjust their local clock.
Hence, $\SM{j}$ describes the \textit{synchronization region} of $M_j$. 

Above definitions are summarized in Table~\ref{tab:symbols}.

\begin{figure}[t]
    \centering
    \includegraphics[width=1\columnwidth]{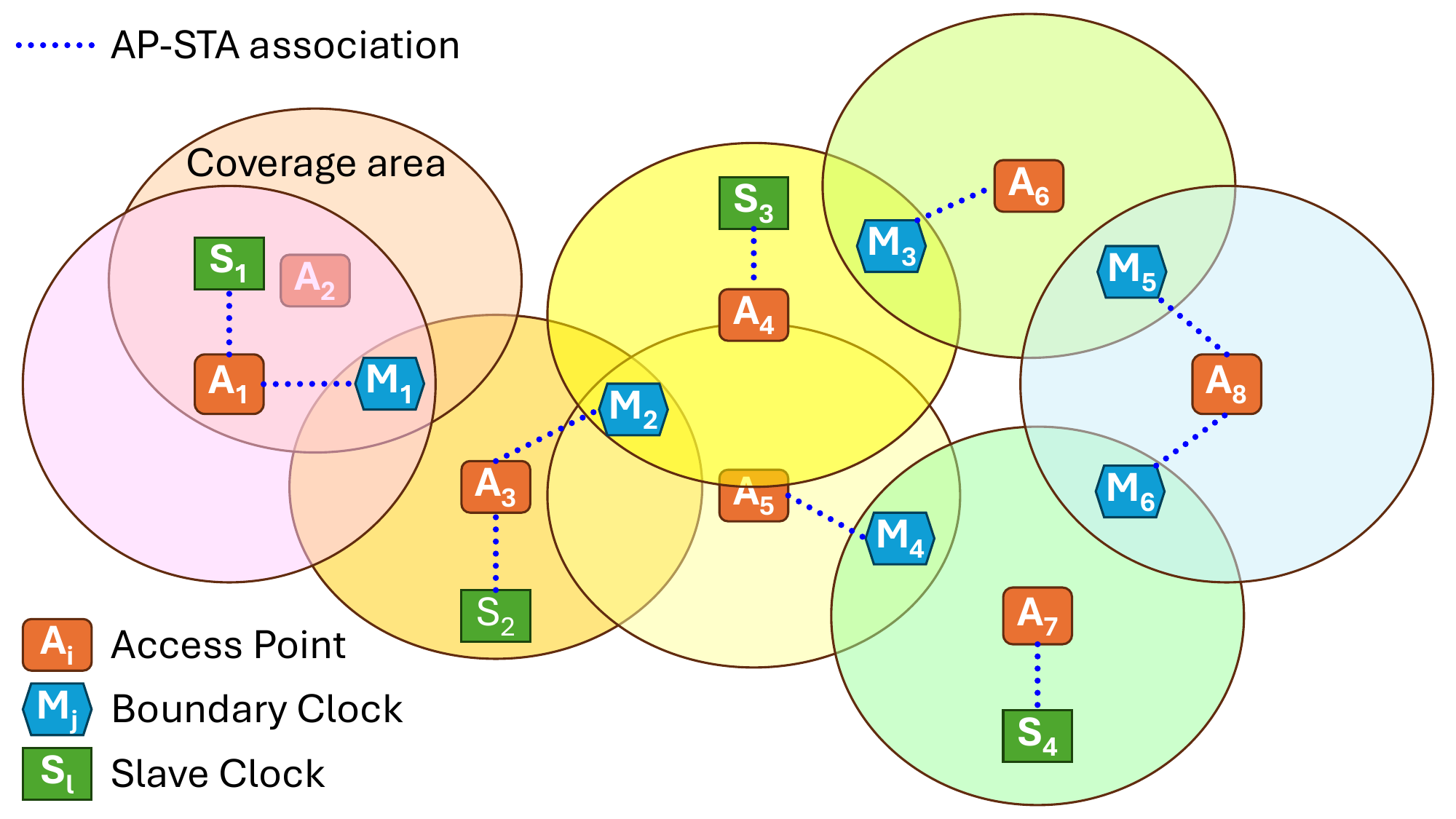}
    \caption{Sample network topology and relevant RBIS nodes.}
    \label{fig:rbis}
\end{figure}

\subsection{Direct Synchronization}
\begin{figure}
    \centering
    \includegraphics[width=1\columnwidth]{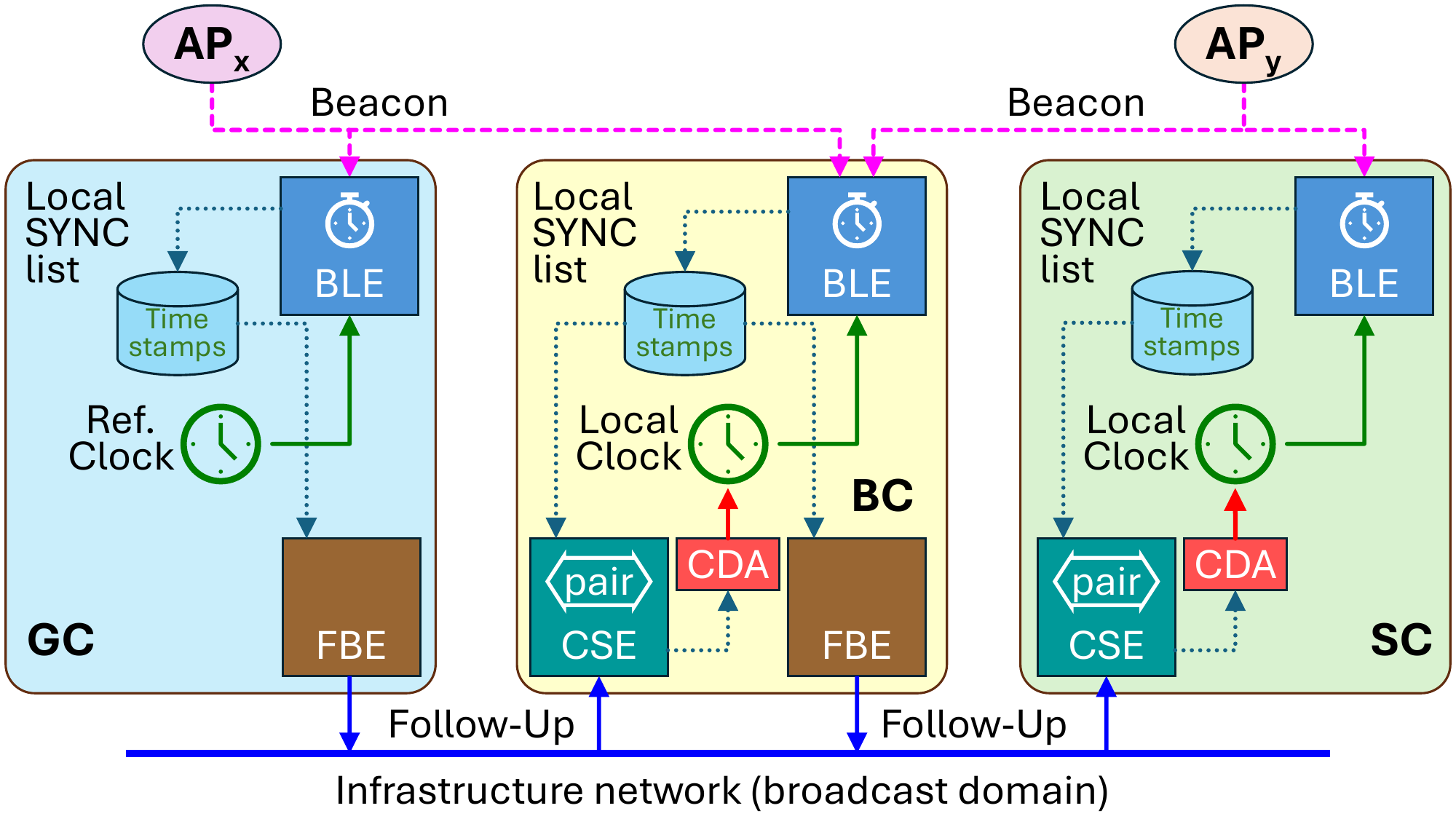}
    \caption{Architecture of RBIS devices (GC, BC, and pure SC).}
    \label{fig:blocks}
\end{figure}

Fig.~\ref{fig:rbis} shows a sample ESS that includes some APs (with the related coverage regions) and several associated TS STAs.
It will be used in the following as a running example to describe the RBIS protocol with extended coverage.
Let us initially consider the GC $M_1$ and the STAs in its synchronization region.
Some operations are performed on every TS STA, while others only take place on either the master or the slave side.
A schematic diagram of the internal architecture of TS-compliant devices is depicted in Fig.~\ref{fig:blocks},
where functions are modeled by means of specific protocol entities.
The simple scenario analyzed in this section only considers the grandmaster clock (on the left)   
and slave clocks (on the right).

\subsubsection{Operations common to both sides}
Like the original RBIS, every TS STA (including $M_1$) captures timestamps on all the beacons it receives from every nearby AP.
Referring to the example in the figure, this means that $M_1$ logs all beacons $\mathfrak{B}^{A_i}_k$ 
sent by $A_i \in \mathcal{A}^{M_1} = \{ A_1, A_2, A_3\}$, and not only those sent by $A_1$ to which it is associated.
Not necessarily APs send beacons at the same rate, although their generation period $T_\mathrm{B}$ is usually set to the default value  $\SI{102.4}{ms}$ (about $\SI{10}{Hz}$).

Beacons acquired by every TS STA $S_l \in \mathcal{S}$ (GC, MCs, and SCs) 
are orderly saved in a local \textit{Sync} list $\mathcal{L}^{S_l}$ as tuples 
$\mathbf{b} = \langle A_i.\mathrm{ID}, \mathfrak{B}^{A_i}_k.\mathrm{TSF}, \mathfrak{B}^{A_i}_k.t \rangle$,
where $A_i.\mathrm{ID}$ identifies the originating AP (by means of its $\SI{6}{B}$ MAC address),
$\mathfrak{B}^{A_i}_k.\mathrm{TSF}$ represents the TSF field in beacon $\mathfrak{B}^{A_i}_k$, and
$\mathfrak{B}^{A_i}_k.t$ is a precise timestamp taken by $S_l$ with its local clock
(that, in the particular case where $S_l$ is the GC $M_1$, coincides with the reference clock to which all STAs in $\mathcal{S}$ will synchronize) on the arrival of $\mathfrak{B}^{A_i}_k$.
Above operations are performed by a specific \textit{beacon logging entity} (BLE),
one instance of which is included in every TS STA.

\subsubsection{Operations on the master clock side}
Every MC $M_j$ (in our example $M_1$, which is the GC) cyclically broadcasts a FUP message $\mathfrak{F}^{M_j}_\ell$ that includes the most recent elements of its local \textit{Sync} list $\mathcal{L}^{M_j}$ (e.g., with a period equal to $T_\mathrm{F}$, details on this will be provided later).
Possibly, a specific multicast MAC address could be reserved for these messages to avoid interrupting STAs that are not involved in clock synchronization (non-TS STAs).
Being targeted to a group MAC address, FUP messages are first sent by $M_j$ to the associated AP ($A_1$, in the example) as unicast, and then broadcast by this AP on its BSS ($\mathrm{BSS}_1$).
Since APs are assumed to be interconnected through Ethernet, every FUP message is also relayed to all the other APs in the ESS ($A_i \in \mathcal{A}$) and, from there, to all the STAs in the related BSSs
(another significant example of broadcast frame transmission in an ESS is given, e.g., by ARP requests).
Therefore, FUP messages reach the entire set $\mathcal{S}$ of STAs.
Above operations are performed by a specific \textit{follow-up broadcasting entity} (FBE),
one instance of which is included in every MC.

By using the above mechanism, all STAs in $\SM{1}$ can synchronize directly to $M_1$.
Looking at the example, besides $S_1$ (that is associated to $A_1$), also $S_2$ (that is associated to $A_3$) can be synchronized.
In fact, both $S_2$ and $M_1$ can hear beacons from $A_3$, and FUP messages sent by $M_1$ reach $S_2$ passing through $A_1$, the Ethernet network, and $A_3$.
More in general, \textit{direct synchronization} of MC $M_j$ and STA $S_l$ (which, with respect to $M_j$, behaves as a SC)
is possible whenever $S_l \in \SM{j}$, and relies on FUP messages generated by $M_j$ and 
beacons sent by any of the APs in $\mathcal{A}^{M_j} \cap \mathcal{A}^{S_l}$.

\subsubsection{Operations on the slave clock side}
All STAs in $\mathcal{S}$ (both those which can behave as MC and those that only support the pure SC role) must implement one instance of the \textit{clock synchronization entity} (CSE), 
which adjusts their local clock based on the time distributed by upstream devices in the time hierarchy.
The CSE, whose operations are described below, is inactive on the GC (root), as this node does not synchronize its local clock this way.

As an example, let us consider SC $S_2$ in Fig.~\ref{fig:rbis} (but the same reasoning applies to MCs as well).
Thanks to the BLE, every beacon $\mathfrak{B}^A_k$ heard by $S_2$,
including those from any $A \in \mathcal{A}^{M_1} \cap \mathcal{A}^{S_2} = \{ A_3 \}$ 
($A_3$ is the only AP seen by both $M_1$ and $S_2$) is saved,
together with its precise arrival time $t$ and TSF, in the local \textit{Sync} list $\mathcal{L}^{S_2}$.
When a FUP message is received (in the example, $\mathfrak{F}^{M_1}_\ell$ sent by $M_1$), the list of tuples it contains,
we denote $\mathcal{R}^{M_1}_\ell$, and the local \textit{Sync} list $\mathcal{L}^{S_2}$ are \textit{paired}.
This means, that these lists are checked looking for entries characterized by the same values for both the ID of the AP and the TSF field (which implies that they refer to exactly the same instance of beacon frame, previously captured on air by $M_1$ and $S_2$, respectively, which is taken as a common \textit{Sync} event).
In formulas, two tuples 
$\mathbf{b}^\mathrm{R} \in \mathcal{R}^{M_j}_\ell$ and $\mathbf{b}^\mathrm{L} \in \mathcal{L}^{S_l}$ match when 
$\mathbf{b}^\mathrm{R}.\mathrm{ID} = \mathbf{b}^\mathrm{L}.\mathrm{ID} \wedge \mathbf{b}^\mathrm{R}.\mathrm{TSF} = \mathbf{b}^\mathrm{L}.\mathrm{TSF}$.
Every matching entry of the pairing process constitutes a \textit{synchronization opportunity} (SYNOP) for $S_2$, 
conceptually conveyed over the \textit{time synchronization link} (TS link) we denote as $M_1 \Rightarrow S_2$.

If at least one match $\langle \mathbf{b}^\mathrm{R}, \mathbf{b}^\mathrm{L} \rangle$ is found, 
pairing is considered successful and the related timestamps (local and remote)
can be used to adjust the clock on $S_2$, aligning it to the reference time from $M_1$.
From a conceptual viewpoint, offset correction amounts to $o = \mathbf{b}^\mathrm{L}.t - \mathbf{b}^\mathrm{R}.t$.
Because of the wireless communication support and since both beacons and FUP messages are broadcast (which implies that no retransmissions can be performed)
there is a non-negligible probability that pairing fails, in which case time synchronization is not carried out.
This probability can be lowered by including information about more than one beacon in the FUP message.
By considering two subsequent FUP messages, rate correction is also possible.
In the case pairing provides more than one match, the most recent tuple (that is, the one with the highest $t$ value) is used for offset correction,
and rate correction becomes (in theory) possible using just one FUP message.

Generally speaking, a suitable CDA should be envisaged to lower the mean synchronization error, avoiding at the same time abrupt changes of the local time (time warps).
CDAs will not be analyzed here, as this aspect is quite peculiar and has already been studied thoroughly in the literature for PTP and NTP, and even for RBIS.
It is worth pointing out that having multiple matches from pairing, i.e., a set of pairs $\{ \langle \mathbf{b}^\mathrm{R}, \mathbf{b}^\mathrm{L} \rangle \}$ that refer to distinct beacons carried in the same FUP message, could be exploited to improve synchronization accuracy and precision.

\section{DOMINO}
\label{sec:DOMINO}
With limited protocol modifications, areas much larger than $\SM{1}$ can be covered, 
potentially including the entire set $\mathcal{S}$.
To do so, master clocks other than the grandmaster (non-GC MCs) are needed, 
which are able to behave as \textit{boundary clocks} (BC) by synchronizing overlapping coverage regions.

\begin{figure}[t]
    \centering
    \includegraphics[width=1\columnwidth]{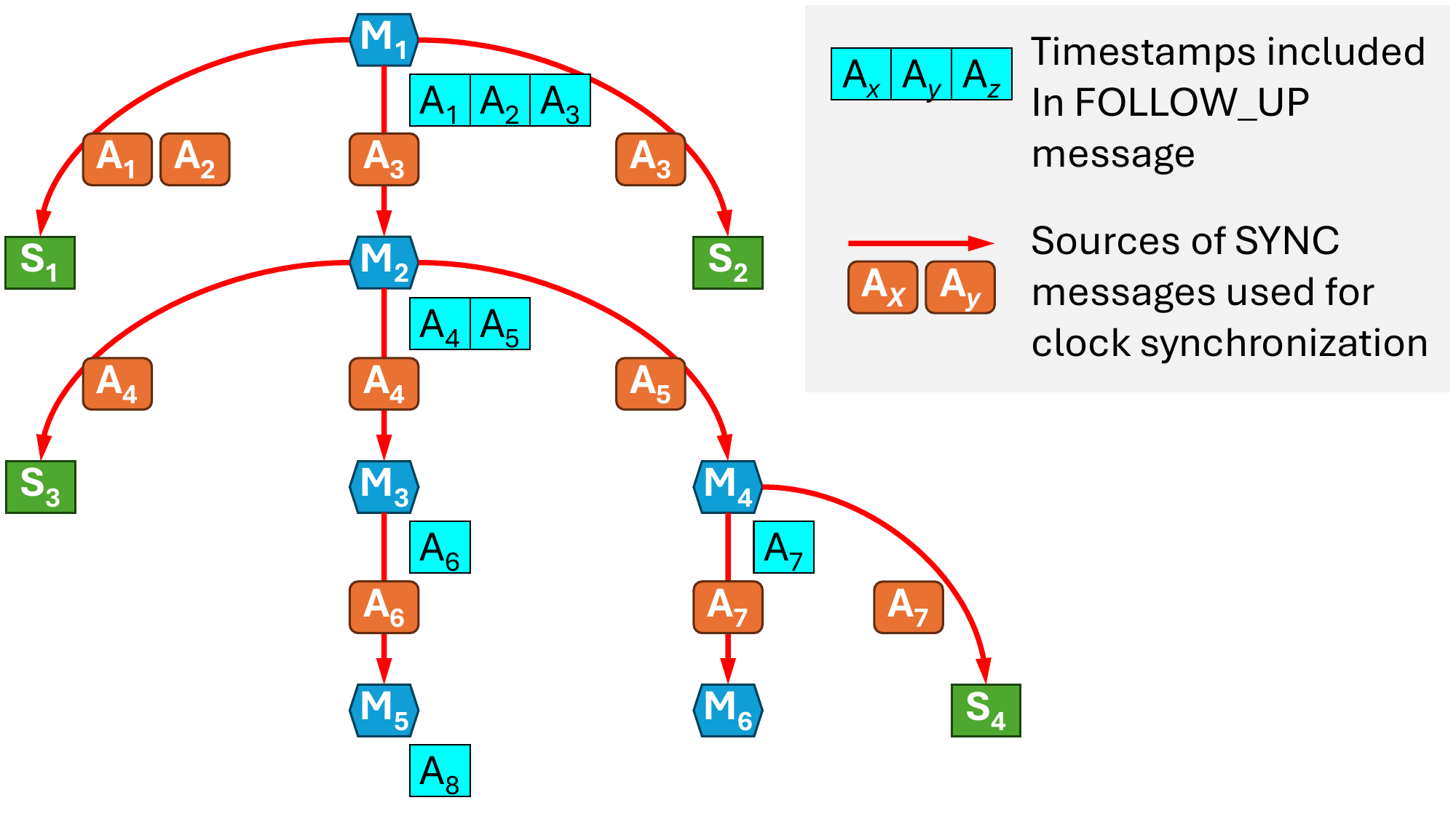}
    \vspace{-4mm}
    \caption{Synchronization tree for the sample RBIS network.}
    \vspace{-4mm}
    \label{fig:tree}
\end{figure}

\subsection{Indirect Synchronization}
Basically, BCs in RBIS with extended coverage are TS STAs that behave as both SC and MC at the same time.
Unlike PTP, where BC functions are implemented by switches 
(which means that master and slave roles are assigned to distinct Ethernet ports), STAs in Wi-Fi do not have separate physical interfaces and all messages pass through the same radio PHY (antenna).
Hence, these roles concern the STA as a whole, which will (virtually) expose both a slave and a master side.
As depicted in the middle part of Fig.~\ref{fig:blocks}, the \textit{slave side} of a BC synchronizes its local clock using FUP messages generated by nearby upstream MCs.
Contextually, its \textit{master side} generates FUP messages (where the included timestamps are taken with the local clock), which propagate time synchronization downstream in $\mathcal{S}$.

Supporting indirect synchronization requires that a plurality of BCs are available in $\mathcal{S}$, deployed in suitable places.
This can be easily achieved by including BC capabilities in every full-function TS STA (FFTS).
Simpler and cheaper reduced-function TS STAs (RFTS) can be also envisaged that only support the slave role.
As an example, let us refer again to the example in Fig.~\ref{fig:rbis}.
The GC $M_1$ can hear beacons from $\mathcal{A}^{M_1} = \{ A_1, A_2, A_3 \}$,  but not from $A_4$.
This means that all STAs in $\mathcal{S} \setminus \SM{1}$ (like $S_3$, which belongs to $\SM{2}$ but not to $\SM{1}$) cannot synchronize with $M_1$ directly. 
However, $M_2$ can hear beacons from both some AP in
$\mathcal{A}^{M_1}$, e.g., $A_3$, as well as from $A_4$ (plus, possibly, other APs as well).
In our case, $\mathcal{A}^{M_2} = \{ A_3, A_4, A_5 \}$.
Then, $M_2$ can act as a BC that synchronizes its local clock to $M_1$ using $\mathfrak{F}^{M_1}$ messages
(in this case $M_2$ behaves as a SC for $M_1$) 
and then broadcasts its own $\mathfrak{F}^{M_2}$ messages to synchronize STAs in $\SM{2} \setminus \SM{1}$,
i.e., all those STAs that can be synchronized by $M_2$ with the exclusion of those that are already synchronized (directly) by $M_1$.
For them, it acts as the MC.

By using the indirect approach above, synchronization in large networks propagates from the GC to every single STA, 
possibly traversing several intermediate BCs,
for example, $M_1 \Rightarrow M_2 \Rightarrow S_3$. 
Overall, this creates a ``synchronization wavefront'' in the ESS that resembles a domino effect (whence the name of the protocol) and may potentially reach BSSs far away from the synchronization region of the GC.
Practically, a synchronization graph can be defined that specifies the relationships between the clocks of the TS STAs, 
but unlike PTP it is not a tree (which means that some mechanism is required to make it loop-free) and may vary over time (sometimes with non-slow dynamics).
For example, the tree in Fig.~\ref{fig:tree} refers to the RBIS setup in Fig.~\ref{fig:rbis} after unnecessary time synchronization links have been cut off.

As for PTP, every time a BC is traversed synchronization accuracy and precision unavoidably worsen.
This means that the synchronization error grows along downstream synchronization paths.
Nevertheless, if suitable offset/rate corrections are performed by the CDA and paths are reasonably short, such additional errors remain acceptably low.
This is not a severe limitation, since DOMINO is not meant to offer the same very high precision as PTP.

\subsection{Avoiding synchronization clashes}
With the above approach, more than one synchronization path may exist between the GC $M_1$ and any target STA $S_l$, which differ for the sequence of traversed BCs (that is, the related chain of TS links).
If multiple FUP messages arrive to $S_l$ for which pairing is successful,
distinct timestamps become available for the same captured beacons, taken by different BCs, all of which constitute valid SYNOPs.
We will refer to this event as a \textit{synchronization clash}, and the final TS links of the involved paths (and the related FUP messages) are said to be clashing.
It must be remarked that having more than one matching entry in the same FUP message does not lead to any clashes.
In this case, all the included timestamps are taken with the same time base (the local clock of the BC that sent that message), even if they refer to beacons generated by different APs.

Referring to the example in Fig.~\ref{fig:rbis}, $S_4$ receives usable FUP messages from both $M_4$ and $M_6$ ($\mathfrak{F}^{M_4}$ and $\mathfrak{F}^{M_6}$, respectively).
In the former case beacons $\mathfrak{B}^{A_7}$ sent by $A_7$ are timestamped using the local clock of $M_4$, while in the latter the clock of $M_6$ is employed.
The related synchronization paths are $M_1 \Rightarrow M_2 \Rightarrow M_4 \Rightarrow S_4$ and
$M_1 \Rightarrow M_2 \Rightarrow M_3 \Rightarrow M_5 \Rightarrow M_6 \Rightarrow S_4$, respectively.
Using both for adjusting the local clock is possible, but not suggested since this would increase jitters.
In fact, after pairing, subsequent remote timestamps are likely to be taken with different clocks, which may deviate in opposite directions from the reference time, 
increasing jitters noticeably (and worsening synchronization precision).
Thus, it is necessary for $S_4$ to decide from what TS link the SYNOPs for adjusting its clock have to be taken.
For the reason stated above about the dynamic nature of the network topology and disturbance, this decision must be made at runtime and not once and for all.
A reasonable solution is to determine which one, among all clashing TS links, ensures the lowest mean error.
The BC that originates the related FUP messages, we denote $M$, is then selected as the \textit{parent clock} (PC) of $S_l$ in the time hierarchy, and only the SYNOPs coming from $M$ will be used for adjusting its local clock. 

\begin{figure}[b]
    \vspace{-0.4cm}
    \centering
    \includegraphics[width=1\columnwidth]{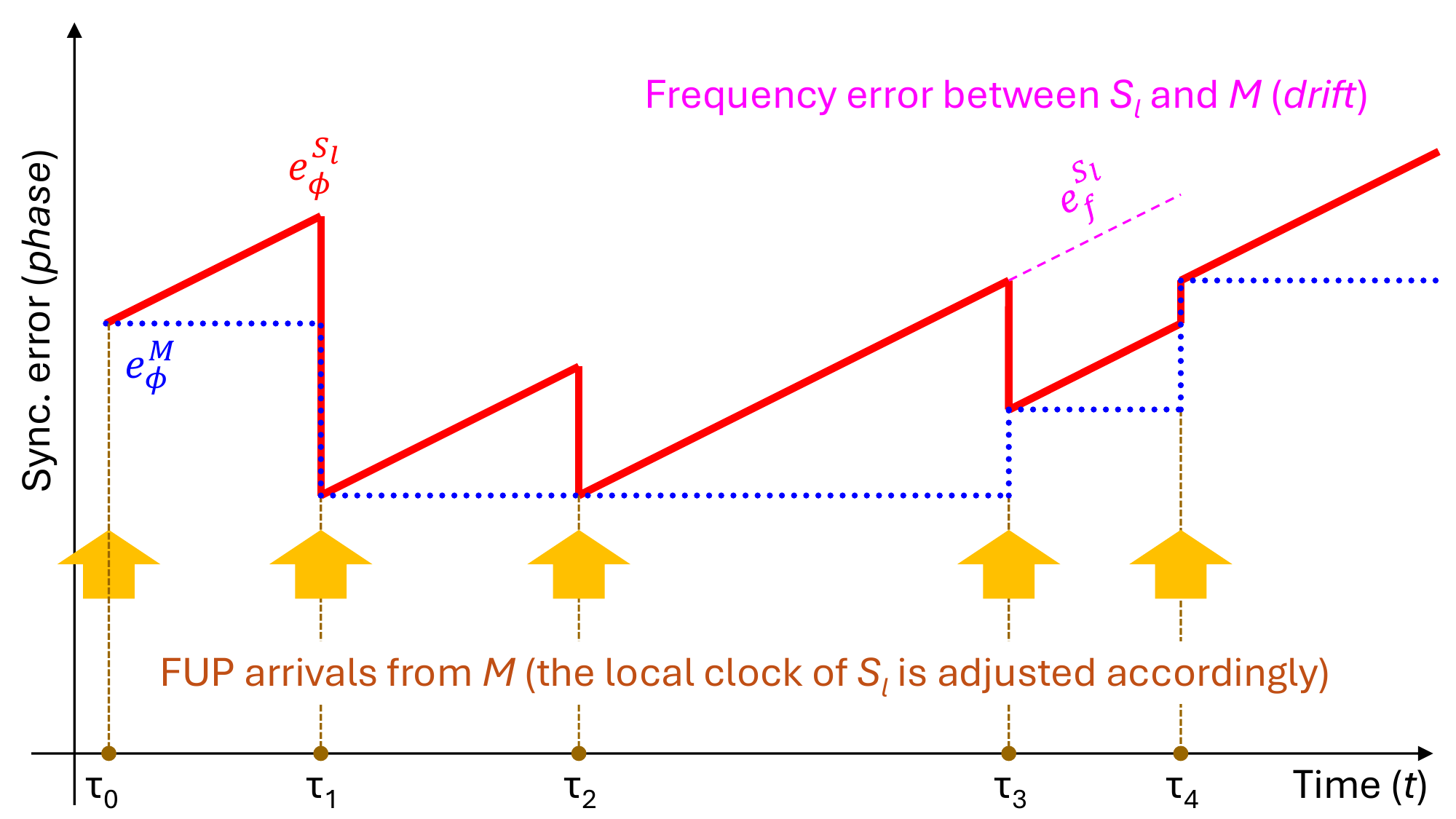}
    \caption{Simple model for approximating the synchronization error of $S_l$.}
    \label{fig:drift}
\end{figure}

Let $T_{\mathfrak{F}^M}$ be the time elapsed from the most recent paired FUP message received from $M$
(which is the last time clock adjustment to $M$ was possible).
Several ways exist for evaluating the error that characterizes a TS link $M \Rightarrow S_l$.
At any given time, we model the synchronization error (skew) $e_\phi^{S_l}$ of $S_l$ as made up of two contributions that depend, respectively, on:
1)~the absolute synchronization error $e_\phi^{M}$ of $M$; and,
2)~the absolute frequency error (oscillator tolerance) $e_f^{S_l}$ with which the clock of $S_l$ drifts away.
A sample plot describing how the clock skew is assumed to change over time (sawtooth wave) is shown in Fig.~\ref{fig:drift}.
As can be seen, the mean synchronization error of $S_l$ when $M$ is selected as its parent clock can be approximated by
\begin{align}
    \label{eq:synerr}
    \overline{e}_\phi^{S_l} = \overline{e}_\phi^{M} + \frac{1}{2} e_f^{S_l} \cdot \overline{T}_{\mathfrak{F}^M},
\end{align}
where $\overline{T}_{\mathfrak{F}^M}$ is the mean intertime between SYNOPs on $M \Rightarrow S_l$, computed by $S_l$ as the exponential moving average (EMA) of the times between the arrivals of subsequent FUP messages from $M$ for which pairing is successful.
Quantity $\overline{e}_\phi^{M}$ is evaluated by $M$ in pretty much the same way as $\overline{e}_\phi^{S_l}$
is computed by $S_l$ (with the sole exception of GC, whose accuracy and precision are defined by other means, e.g., starting from the quality of either its quartz or the external time source it relies on) and is conveyed as a field $\mathfrak{F}^M_{\ell}\!.e$ inside its FUP messages.
Instead, parameter $e_f^{S_l}$ depends on the stability of the local clock of $S_l$ after rate correction (e.g., against temperature variations) 
and can be estimated for every specific device 
given its type and environmental conditions.

Although the above model is simplistic, it satisfactorily describes the accuracy of clock adjustments.
Better models, including details about error calculation, are part of future activities.
The synchronization error is evaluated by $S_l$ for all the clashing TS links using \eqref{eq:synerr}.
Then, the BC that achieves the smallest error is selected as PC.
This corresponds to pruning the synchronization graph in such a way to turn it into a tree
(every TS STA chooses a single parent).

\subsection{Parent clock selection}
Operating conditions of Wi-Fi networks keep varying continuously over time as:
1) mobile STAs are likely to change their position;
2) obstacles with reflective or absorbing surfaces do the same;
3) interference (traffic) generated by nearby STAs also varies (and is often bursty in nature); and,
4) especially in industrial environments, sources of electromagnetic noise repeatedly turn on and off (according to the operating cycles of high-power machinery).
This means that the CSP must reconstruct the synchronization tree dynamically. 

The mechanism every TS STA $S_l$ uses to select its parent clock $M_p$ operates as follows.
The value of $p$ is kept in a local variable, which is initialized to NULL at startup to mean that the STA is \textit{unsynchronized} (the local clock is not adjusted).
Whenever $S_l$ receives a FUP message $\mathfrak{F}^{M_j}_\ell$ and pairing succeeds  
(which means that the sender $M_j$ could be potentially selected as $M_p$),
information about the \textit{clock quality} of the TS link $M_j \Rightarrow S_l$ is saved as a tuple 
$
\mathbf{c}_j = \langle M_j.\mathrm{ID}, e, \tau, \overline{T} \rangle
$
in a locally maintained dynamic data structure $\mathcal{C}^{S_l}$, 
whose entries are uniquely identified by $M_j.\mathrm{ID}$ (which behaves as the primary key).
In particular, $e$ is taken from the message (field $\mathfrak{F}^{M_j}_{\ell}\!.e$)
while $\tau$ represents its arrival time $\tau_\ell^{M_j}$.

If no entry exists for $M_j$ in $\mathcal{C}^{S_l}$, one is \textit{created} and initialized by setting $\overline{T}$ to undefined.
In this case, $\overline{e}_{\phi}^{M_j \Rightarrow S_l}$ is conventionally set to $\infty$, 
since no information is available yet about the mean intertime between SYNOPs over $M_j \Rightarrow S_l$.

Otherwise, if $M_j$ is already known, the related entry is \textit{updated} by computing $\overline{T}$ 
as follows:
if the related value in $\mathbf{c}_j$ is still undefined (this happens the second time a FUP message from $M_j$ is successfully paired on $S_l$), then it is set equal to
\begin{align}
    \label{eq:avgtimeu}
    \overline{T} = \beta ( \tau - \tau' ) + T_0,
\end{align}
where $\beta \geq 1$ (e.g., $2$) and $T_0 \geq 0$ (e.g., $\SI{1}{s}$) are suitable multiplicative and additive inertial coefficients, respectively, which prevent the current PC from being changed too early (by waiting until the new TS link has settled).
Instead, if $\overline{T}$ in $\mathbf{c}_j$ is already defined, the recursive equation 
\begin{align}
    \label{eq:avgtime}
    \overline{T} = \alpha ( \tau - \tau' ) + (1-\alpha) \overline{T}',
\end{align}
is used, where 
$\alpha$ is a smoothing factor (e.g., $0.125$) and the prime symbol denotes the previous value.
When the entry already exists, 
the synchronization error $\overline{e}_{\phi}^{M_j \Rightarrow S_l}$ can be estimated through \eqref{eq:synerr} by setting $\overline{T}_{\mathfrak{F}^{M_j}}$ equal to $\mathbf{c}_j.\overline{T}$.

To prevent the size of the $\mathcal{C}^{S_l}$ data structure from growing excessively, 
if no FUP messages that can be paired are received from a certain $M_j$ for a given time $T_\mathrm{pcl}$ (parent clock lifetime, e.g., one minute), the corresponding entry is \textit{deleted}.

Whenever a change is observed in $\mathcal{C}^{S_l}$ (creation, update, or removal of an entry),
$S_l$ determines its new \textit{best potential} PC, denoted $M_{b}$, as
$b = \mathrm{arg} \min_j \{ e_{\phi}^{M_j \Rightarrow S_l} | \mathbf{c}_j \in \mathcal{C}^{S_l} \}$.
To prevent the PC from being changed too often (which would worsen precision), hysteresis can be exploited:
the current PC $M_p$ is changed to $M_b$ only when doing so is clearly advantageous, 
in spite of statistical fluctuations due to the randomness of the wireless medium.
In formulas, $p$ is set equal to $b$ only if
$e_{\phi}^{M_b \Rightarrow S_l} < \alpha e_{\phi}^{M_p \Rightarrow S_l}$,
where $\alpha < 1$ (e.g., $\alpha = 0.875$).

If the STA is currently \textit{unsynchronized} (which implies that $\mathcal{C}^{S_l}$ is empty),
then as soon as a FUP message $\mathfrak{F}^{M_j}$ is received that is successfully paired, 
$p$ is set equal to $j$ (that is, $M_j$ becomes immediately the new PC) and the TS STA is turned to the \textit{synchronized} state.

\subsection{Dynamic selection of the reference clock}
As seen before, synchronization starts at the grandmaster clock and propagates along paths that, in the indirect case, traverse one or more boundary clocks.
Because of network dynamics, a different BMC algorithm should be used with respect to PTP, which relies on FUP messages.
In particular, every TS STA that can behave as a GC (there can be more than one in the network, that in general coincide with BCs) defines its \textit{local clock quality} ($Q_{\mathrm{L}}$).
To this purpose, the same descriptors used by the PTP BMCA and included in \textit{Announce} messages can be adopted, which embed, in the given order, the following properties:
priority 1, clock quality class, accuracy, and variance, priority 2, and clock identity (that corresponds to the MAC of the related TS STA).
The lower the related values, the higher the quality of the clock, which is used in turn to select the GC.
When configuring the protocol, care must be taken so that no two BCs may have the same $Q_{\mathrm{L}}$
(this property always holds when MAC addresses are used as identities).
For pure SCs, $Q_{\mathrm{L}}$ is set conventionally to $\infty$.

Besides $Q_{\mathrm{L}}$, every BC also defines a similar descriptor, termed \textit{reference clock quality} ($Q_{\mathrm{R}}$), which reflects the quality of the reference clock source it is currently using.
At startup, when the BC is not synchronized yet, $Q_{\mathrm{R}}$ is set equal to $Q_{\mathrm{L}}$, i.e., it assumes to be the GC.
In this case, it considers its local clock as the reference clock of the network, and does not adjust it through the CSE.
Every BC includes its current $Q_{\mathrm{R}}$ in outgoing FUP messages, exploiting a specific \textit{source clock quality} (SQ) field.
This permits to characterize the \textit{native} quality of the time base used to acquire its timestamps.
Whenever a FUP message is received that can be successfully paired, the enclosed SQ field is checked against $Q_{\mathrm{R}}$.
If SQ is better (i.e., lower) than $Q_{\mathrm{R}}$, $Q_{\mathrm{R}}$ is set equal to SQ, to mean that a better reference clock has been found in the network (and the PC is possibly changed).
In this way, after some time has elapsed from the moment the current GC (that is unique network-wide) was switched on, all TS STAs eventually set their $Q_{\mathrm{R}}$ to the GC's $Q_{\mathrm{L}}$ and are synchronized (possibly across several hops) to its clock.

To support dynamic GC selection, the procedure to determine the PC must be modified as follows.
First, a new field $q$ is added to tuples $\mathbf{c}_j$, which stores the SQ descriptor in incoming paired FUP messages.
Second, 
an entry can be created/updated in $\mathcal{C}^{S_l}$ only if the SQ of the incoming paired FUP message is better than $Q_{\mathrm{R}}$ (adjusting the local clock with a worse quality one is pointless). 
Otherwise the FUP message is ignored.
Third,
the entry in $\mathcal{C}^{S_l}$ with the lowest $q$ value is selected as the new PC
(change is immediate if $Q_{\mathrm{R}}$ improves).
The computed error $e_{\phi}^{M_j \Rightarrow S_l}$ is only considered for discriminating among  entries with the same $q$ value.
Fourth, $Q_{\mathrm{R}}$ is set equal to the $q$ value of the entry $\mathbf{c}_p$ selected as the PC.
Because of the strict ordering enforced on clock quality, above procedure prevents loops in the synchronization graph.
When operations of the wireless BMC algorithm (WBMCA) settle, all active TS links form a proper synchronization tree, 
whose root coincides with the current GC.

Lifetime $T_{\mathrm{pcl}}$ set on $\mathcal{C}^{M_j}$ entries also serves to deal with those cases where the current GC ceases operation (e.g., it is turned off or moves out of the area related to $\mathcal{S}$).
If, at any given time, the  $Q_{\mathrm{R}}$ of all nearby active BCs of $M_j$ are worse than its $Q_{\mathrm{L}}$,
then after a time $T_{\mathrm{pcl}}$ the structure $\mathcal{C}^{M_j}$ becomes empty
and $M_j$ assumes (again) to be the GC, setting its $Q_{\mathrm{R}}$ equal to $Q_{\mathrm{L}}$.
In the case a better GC becomes subsequently available in the network, a synchronization path will eventually appear directed from GC to one of the neighbor BCs of $M_j$, which creates an entry in $\mathcal{C}^{M_j}$ that is taken as its new PC.

\section{Conclusions}
\label{sec:concl}
Precise clock synchronization protocols are being more and more exploited in real networks for two reasons: 
first, they permit real-time distributed applications to co-ordinate their actions, and second, they enable network equipment to support deterministic communication (see, e.g., TSN). 
While standardized CSP solutions exist for wired networks like Ethernet, this is not the case when wireless local area networks are taken into account, the most popular being with no doubts Wi-Fi.
While the same PTP mechanisms could be satisfactorily applied also in that case, as a matter of fact very few hardware exists  (if any) that supports this in a standard manner.

RBIS is a simple protocol that can be implemented in software in Wi-Fi end devices (STAs).
The ability to take precise timestamps on the wireless adapter permits to lower  the synchronization error tangibly, but is not strictly required.
Interestingly, no modifications at all are required on APs by RBIS, which means that the existing network infrastructure can be left untouched (in theory, at least).
The only serious drawback of RBIS is its limited coverage, which can not stretch beyond the related BSS.

In this paper, a version of RBIS with extended coverage is described, we term DOMINO.
Unlike similar solutions in the literature, DOMINO is explicitly conceived to optimize synchronization quality in spite of phenomena that typically affect wireless networks, that is, mobility of devices and disturbance.
The former issue demands for dynamic solutions, which can rebuild the synchronization hierarchy (tree) on the fly when the position of STAs keep changing.
The second issue, instead, has to do with the fact that messages used by RBIS (both FUP and beacons) are sent as broadcast, and hence they may suffer from severe losses.
The CSP, in this case, should be able to prefer longer multi-hop synchronization paths that traverse several BCs over shorter (or direct) ones, if the mean time between synchronization opportunities on the latter is too long to ensure proper clock adjustments (because, e.g., the majority of FUP messages and, especially, beacons are lost, due to either attenuation or localized interference). 

A formal proof of correctness for the protocol and performance evaluation
(accuracy and settling times) under severe disturbance and device mobility are part of our future work.

\bibliographystyle{IEEEtran}
\bibliography{bibliography}

\end{document}